# High-performance monolayer MoS$_2$ field-effect transistor with large-scale nitrogen-doped graphene electrodes for Ohmic contact

Dongjea Seo,[1] Dong Yun Lee,[2] Junyoung Kwon,[1] Jea Jung Lee,[3] Takashi Taniguchi,[4] Kenji Watanabe,[4] Gwan-Hyoung Lee,[5] Keun Soo Kim,[2] James Hone,[6] Young Duck Kim,[3,a)] and Heon-Jin Choi[1,a)]

[1]Department of Materials Science and Engineering, Yonsei University, Seoul 03722, Republic of Korea

[2]Department of Physics & Graphene Research Institute, Sejong University, Seoul 05006, Republic of Korea

[3]Department of Physics, Kyung Hee University, Seoul 02447, Republic of Korea

[4]National Institute for Materials Science, 1-1 Namiki, Tsukuba 305-0044, Japan

[5]Department of Materials Science and Engineering, Seoul National University, Seoul 08826, Republic of Korea

[6]Department of Mechanical Engineering, Columbia University, New York, New York 10027, USA

[a)] Correspondence to: ydk@khu.ac.kr (+82-10-3796-8916), hjc@yonsei.ac.kr (+82-10-5232-2235)

A finite Schottky barrier and large contact resistance between monolayer MoS$_2$ and electrodes are the major bottlenecks in developing high-performance field-effect transistors (FETs) that hinder the study of intrinsic quantum behaviors such as valley-spin transport at low temperature. A gate-tunable graphene electrode platform has been developed to improve the performance of MoS$_2$ FETs. However, intrinsic misalignment between the work function of pristine graphene and the conduction band of MoS$_2$ results in a large threshold voltage for the FETs, because of which Ohmic contact behaviors are observed only at very high gate voltages and carrier concentrations (~10$^{13}$ cm$^{-2}$). Here, we present high-performance monolayer MoS$_2$ FETs with Ohmic contact at a modest gate voltage by using a chemical-vapor-deposited (CVD) nitrogen-doped graphene with a high intrinsic electron carrier density. The CVD nitrogen-doped graphene and monolayer MoS$_2$ hybrid FETs platform exhibited a large negative shifted threshold voltage of -54.2 V and barrier-free Ohmic contact under zero gate voltage. Transparent contact by nitrogen-doped graphene led to a 214% enhancement in the on-current and a four-fold improvement in the field-effect carrier mobility of monolayer MoS$_2$ FETs compared with those of a pristine graphene electrode platform. The transport measurements, as well as Raman and X-ray photoelectron spectroscopy analyses before and after thermal annealing, reveal that the atomic C-N bonding in the CVD nitrogen-doped graphene is responsible for the dominant effects of electron doping. Large-scale nitrogen-doped graphene electrodes provide a promising device platform for the development of high-performance devices and the study of unique quantum behaviors.

Two-dimensional (2D) transition metal dichalcogenides (TMDCs) have unique electrical and optical properties. Of them, monolayer molybdenum disulfide ($MoS_2$) is theoretically expected to be used as material for a replacement channel for post-silicon electronics, such as a field-effect transistors (FETs), owing to its direct bandgap (1.9 eV), high electron mobility, high transconductance, and excellent on/off ratio (> $10^8$).[1–3] In addition, the conduction and valence band edges of monolayer $MoS_2$ can indicate valley degrees of freedom for next-generation optoelectronics.[4,5] Because of the large valley separation in the momentum space, an inversion symmetry breaking with strong spin-orbit coupling leads to spin-valley physics that enables the manipulation of the spin and valley in monolayer $MoS_2$. Despite these potential advantages, there is a major bottleneck in developing high-performance $MoS_2$-based devices and studying their exotic spin-valley quantum physics because the charge transport in $MoS_2$ devices is largely dominated by poor electrical contact. Due to Fermi level pinning at the metal–$MoS_2$ interface, most metals form the Schottky contact that interferes with efficient charge carrier injection and extraction, and this often limits the study of the intrinsic transport properties of $MoS_2$.

Past attempts to solve the problem of poor electrical contact have featured the selection of the most suitable metal for $MoS_2$ to overcome Schottky contact,[6–8] but have achieved limited success. In the relevant approaches, achieving the lowest contact resistance using low-work function scandium contact[9] has been shown to lower the Schottky barrier height (SBH) to ~30 meV. The large contact resistance is the result of Fermi level pinning in $MoS_2$ near the conduction band due to the charge neutral level position[10] as well as sulfur vacancy or the defect level.[11,12] In addition, experimental and computational studies suggest that absorbed contamination and damage due to kinetic energy transfer during metal deposition can result in Fermi level pinning and an increase in contact resistance.[13,14] Other approaches to contact engineering, including the use of edge contact,[15,16] phase engineering,[17,18] thermal annealing,[19] and selective etching,[20,21] have limitations for approaching to Ohmic contact. To prevent Fermi level pinning, field-effect transistors based on graphene–$MoS_2$ van der Waals heterojunctions exhibit high mobility and on/off ratio.[1,22] The Fermi level of graphene can be easily changed by using the gate voltage to ensure a match with the conduction band of $MoS_2$.[1,22,23] The use of graphene electrode platforms is the best-known strategy at present for developing high-performance monolayer $MoS_2$ FETs and observations of quantum transports such as the quantum Hall effect and the Shubnikov-de Haas oscillation.[1] However, reliable Ohmic contact with monolayer $MoS_2$ FETs at low temperature has been demonstrated only at very high carrier density ($n$ ~1 × $10^{13}$ $cm^{-2}$) due to the finite difference in the work function between graphene and monolayer MoS$_2$. In a previous study,[24] we used monolayer hBN/Co contact to achieve Ohmic contact under a modest carrier density regime ($n < 10^{13}$ cm$^{-2}$), but this strategy has fundamental limitations in terms of device stability and large-scale fabrication.

In this study, we propose a large-scale highly electron-doped graphene contact platform of monolayer MoS$_2$ for reliable Ohmic contacts and Fermi level alignment. We demonstrate monolayer MoS$_2$ FETs using nitrogen-doped graphene (NGr) with barrier-free Ohmic contact, instead of the pristine graphene contact platforms that were previously used. The highly electron-doped graphene leads to low contact resistance, negative threshold voltage, and high performance of the MoS$_2$ device. Further, we performed electrical transport, Raman spectroscopy and X-ray photoelectron spectroscopy (XPS) before and after thermal annealing to determine the reason for the dominant effects of electron doping in NGr.

Mechanical exfoliated monolayer MoS$_2$ [atomic structure shown in Fig. 1(a)] was characterized by Raman and photoluminescence (PL) spectroscopies. Figure 1(b) shows the Raman spectrum of monolayer MoS$_2$, which exhibited two major peaks corresponding to an in-plane (E$_{2g}$) mode at 384.72 cm$^{-1}$ and an out-of-plane (A$_{1g}$) mode at 403.62 cm$^{-1}$ at an excitation laser wavelength of 532 nm. The Raman peak difference (~19 cm$^{-1}$) can be used to reliably identify monolayer MoS$_2$. Furthermore, we observed PL peaks at 1.86 eV (666 nm) and 1.99 eV (623 nm) corresponding to the A1 and B1 excitons of monolayer MoS$_2$ as shown in Fig. 1(c).

We demonstrated the controlled growth of highly electron-doped NGr on Cu foil by CVD method at 1000 °C with pyridine (C$_5$H$_5$N) liquid precursor, which is a nitrogen-contained organic molecule, instead of the ammonia (NH$_3$) gas precursor. Because a high concentration of ammonia gas precursor leads to the oxidation of Cu foil during CVD graphene growth, and results in an insufficient nitrogen doping and high defect density in NGr.[25] On the contrary, the liquid pyridine (C$_5$H$_5$N) precursor as the source of both carbon and nitrogen enables an efficient nitrogen doping and stable growth of monolayer NGr over a large area with reduced defect density.[26,27] The detailed method of NGr synthesis is described in Supporting Information 1.

Figure 1(d) shows schematics of three types of possible substitutional nitrogen dopant in NGr: graphitic, pyridinic, and pyrrolic. The incorporation of substitutional nitrogen dopant into the carbon lattice is a direct way to control the electronic structure of graphene.[28,29] The graphitic−nitrogen in NGr induces n-type conductivity because an electron participates in the π bond and the fifth electron forms a partial π* bond in the conduction band,[30] whereas the pyridinic- and pyrrolic-nitrogen in graphene form p-type dopants.[31]

To characterize the effect of the doping of nitrogen in graphene, we performed Raman spectroscopy on NGr and pristine graphene. In Fig. 1(e), the pristine graphene (red line) shows two intense Raman peaks corresponding to G (~1588.52 cm$^{-1}$) and 2D (~2684.75 cm$^{-1}$) at an intensity ratio of $I_{2D}/I_G$ = 2.55.[32] In the NGr (black line), the slightly blue-shifted G peak (~1596.68 cm$^{-1}$) of NGr was observed compared with the G peak (~1588.52 cm$^{-1}$) of pristine graphene. There are several reasons for the Raman peak shift, including the effects of both doping and strain.[33] Upon electron doping by nitrogen, the carrier concentration based on the G peak shift was approximately $3.5 \pm 0.2 \times 10^{12}$ cm$^{-2}$. Moreover, a strong D peak at ~1351.37 cm$^{-1}$, a D′ peak at ~1630.88 cm$^{-1}$, and a combination mode ~2955.95 cm$^{-1}$ (D+D′ peak) appeared in NGr. They were activated by such defects as substitution heteroatom, vacancies, and grain boundaries typically observed in NGr.[33-35] Figure 1(f) clearly distinguishes between G and D′ peaks of highly electron-doped NGr compared with those of pristine graphene. The shifts in the G-peak and D′-peak were due to charge distribution, which is indirect evidence of the doping effect of nitrogen (for additional information, see Supporting Information S2).

To further investigate the effect of the doping of NGr, we measured electrical transport under a constant source–drain voltage of 10 mV and observed a charge neutrality point (CNP) at -49.2 V as shown in Fig. 1(g), which corresponded to a carrier density of $3.72 \times 10^{12}$ cm$^{-2}$. This carrier density is consistent with the G peak shift of NGr. The NGr FETs on SiO$_2$ had an electron mobility ($\mu_e$) of ~420 cm$^2$/Vs at $V_g$ = +80 V and hole mobility ($\mu_h$) of ~550 cm$^2$/Vs at $V_g$ = -80 V at room temperature (Supporting Information S3). We confirmed the level of nitrogen doping and high mobility of NGr using previous studies.[26,36] This result shows that the properties of NGr were controlled successfully via CVD method.

Figure 2(a) shows the schematic of the fabrication process of MoS$_2$ FETs with large-scale CVD-grown NGr electrodes. The details are presented in Supporting Information S1. The optical microscopic image of the MoS$_2$ FETs device with NGr is shown in Fig. 2(b), where the NGr electrodes (dashed black line) and monolayer MoS$_2$ (red dashed line) are marked. Figure 2(c) shows the schematic of the final MoS$_2$ FETs device with NGr electrodes encapsulating only the top hexagonal boron nitride (hBN). Figure 2(d) shows the cross-sectional view of the structure of the monolayer MoS$_2$ FETs with NGr electrodes.

We investigated the electrical transport properties of monolayer MoS$_2$ FETs with pristine graphene electrodes as a reference and compared the results with those of the highly doped NGr electrode. Figure 3(a) shows the linear and semi-log plots of the two-probe transfer characteristics of MoS$_2$ FETs obtained at a drain voltage ($V_{ds}$) of 100–400 mV with a 100 mV step. The transfer characteristics of monolayer MoS$_2$ FETs with pristine graphene electrodes exhibited typical n-type behaviors with a threshold voltage ($V_{th}$) of +39 V, where $V_{th}$ was extracted as the gate voltage axis intercept of the linear extrapolation in the linear region in the range of $V_g$ from +60 V to +79 V. The field-effect mobility of MoS$_2$ with pristine graphene electrodes was µ = ~ 2.21 cm$^2$/Vs, which was obtained from $\mu = \frac{L}{wC_{ox}V_{ds}} \frac{dI_{ds}}{dV_g}$, where $\frac{dI_{ds}}{dV_g}$ is transconductance, $L$ (8 µm) and $W$ (9 µm) are the channel length and width, respectively, $C_{ox}$ is the back-gate capacitance per unit area (1.21×10$^{-8}$ F/cm$^2$ for 285-nm-thick SiO$_2$), $I_{ds}$ is the drain current, and $V_g$ is the gate voltage. Figure 3(b) shows the two-probe output curves of the monolayer MoS$_2$ FETs with pristine graphene electrodes depending on the gate voltage from -80 to +80 V, and they indicate the characteristics of the linear output and modest $I_{ds}$ at a high $V_g$. A relatively small $I_{ds}$ for the pristine graphene electrodes for monolayer MoS$_2$ is due to the high contact resistance and a finite SBH between pristine graphene (4.5 eV) and MoS$_2$ (4.15 eV),[37] as shown in Fig. 3(c).

We also performed the transfer- and output-curve characteristics of the monolayer MoS$_2$ FETs with highly doped NGr electrodes. Figure 3(d) shows the linear and semi-log plots of the transfer characteristics of MoS$_2$ FETs with NGr electrodes obtained at range of drain voltage ($V_{ds}$) of 100–400 mV in steps of 100 mV. The $V_{th}$ of NGr electrodes to the monolayer MoS$_2$ FETs was -54.2 V while that of the pristine graphene electrode platform was +39 V. The negative shift in $V_{th}$ of the MoS$_2$ FETs indicates a higher level of electron doping of NGr for carrier transportation at the NGr/MoS$_2$ interface than at the pristine graphene/MoS$_2$ interface as well as a lowering of SBH.

In the MoS$_2$ FETs with NGr electrodes, the on-current improved to 214% compared with that of pristine graphene contact at the same source/drain bias at $V_g$ = +80 V, where the ratio of the on/off current was approximately 10$^6$, similar to previous results.[1,38,39] As shown in Fig. 3(d) ($L$ – 9.3 µm and $W$ – 11.2 µm), the field-effect mobility of the device in contact with NGr was 8.76 cm$^2$/Vs, which was four times that of the device in contact with pristine graphene (2.21 cm$^2$/Vs). Figure 3(e) shows the output curve of MoS$_2$ FETs with varying $V_g$ from -80 V to 80 V. The output curve was linear when both pristine graphene and NGr contact were used for MoS$_2$ at room temperature. Even in the presence of barrier height, the thermal energy at room temperature (26 meV) provided enough energy for the charge to pass over. However, in the output curve at 77 K, as shown in Supporting Information S5, MoS$_2$ with NGr showed the linear Ohmic contact behavior in the gate range from -80 V to +80 V, and in case of pristine graphene contact showed non-linear Schottky contact behavior even at large gate voltage ($V_g$ = +100 V). We can estimate the Fermi level of NGr based on the carrier density (n = 3.72 × 10$^{12}$ cm$^{-2}$, NGr at $V_{CNP}$ = -49.2 V) under zero gate voltage, $E_F = \hbar v_F \sqrt{\pi n}$, where, $v_F$ and $n$ are Fermi velocity and carrier density, respectively. Consequently, the Fermi energy level of NGr can be inferred to be 239 meV higher than that of pristine graphene (4.5 eV).[37] Figure 3(f) shows the corresponding band alignment between NGr and MoS$_2$.

The four-fold improvement in the mobility of monolayer MoS$_2$ FETs with NGr electrodes can be attributed to the barrier-free Ohmic contact for efficient carrier injection and extraction, as shown in Fig. 3(f). Regarding the threshold voltage and SBH, we previously reported a correlation between $V_{th}$ and SBH by adjusting the Fermi level of graphene contact to monolayer MoS$_2$ using dual top and bottom gates.[40] When the Femi level of graphene approached the conduction band of monolayer MoS$_2$, we observed that a reduction in SBH led to a higher conductance as well as a negative shift in $V_{th}$. Based on previous reports, the large negative shift of NGr electrodes contact platform is due to the high Fermi level of NGr compared with the conduction band of monolayer MoS$_2$, which led to Ohmic contact at room temperature and low temperature. The value of $V_{th}$ of the monolayer MoS$_2$ FETs with NGr electrodes was -59.2 *V* while that of the pristine graphene electrode platform was +39 *V*, which was in good agreement with barrier-free Ohmic contact of NGr contact at low temperature. We observed a negative shift in $V_{th}$ and Ohmic contact behaviors from several NGr contact platforms MoS$_2$ (Supporting Information S4).

To determine whether the electrical properties of the MoS$_2$ device were influenced by substitutional nitrogen dopant in NGr, we investigated the electrical transport properties of monolayer MoS$_2$ with NGr contact before and after thermal annealing at 360 °C for 30 min in a vacuum (~ 10$^{-6}$ Torr). The thermal annealing process (360 °C) in a vacuum is the typical process for removing the polymer residue of van der Waals heterostructure fabrication, and do not damage MoS$_2$ with hBN encapsulation.[40,41] Due to vacancy defects in the internal structure of graphene, the substitutional nitrogen dopant in NGr was thermodynamically unstable. The annealing process was carried out to desorb only the substitutional nitrogen dopant in the NGr.

To prove the possibility of degradation of MoS$_2$ (oxidation or sulfur vacancy), the Raman and PL spectroscopes of the monolayer MoS$_2$ were obtained before and after annealing (Supporting Information S6).

The shift and intensity of the Raman peak of MoS$_2$ were almost negligibly small. The PL of MoS$_2$ showed a red shift of 0.02 eV at the peak after annealing due to the effect of strain. The full-width at half maximum (FWHM) of sharp PL peaks of 93 meV (before annealing) and 86 meV (after annealing) of A-exciton also indicate that the hBN-encapsulated MoS$_2$ retained its crystal structure without degradation (Fig. S6).

For accurate analysis of the composition of NGr, we analyzed its binding state and chemical composition before and after annealing by X-ray photoelectron spectroscopy (XPS). Figure 4(a) shows the C 1s XPS spectra of NGr before and after annealing. Before annealing, we observed the strongest peak at ~285 eV (C-C) and the weak peak at ~288 eV (C-N),[42] which originated from the substitution of nitrogen. Following annealing, the only sharp peak at around 285 eV corresponds to the sp2 carbon with C-C bonds. The N 1s XPS spectrum [Fig. 4(b)] featured the formation of pyridinic (397.3 eV), pyrrolic (400.2 eV), and graphitic (401.7 eV) N structures before annealing. This shows that graphitic N atoms were dominant, which can lead to a strong n-type doping effect. However, after annealing, N-C bonding in the nitrogen binding states (N 1s) was not observed. This is clear evidence that the substituted nitrogen was desorbed from NGr by thermal annealing.

To determine the effect of annealing on the reduction in the substitutional nitrogen dopant in NGr, we measured the electrical transport of NGr [Fig. 4(c)] and monolayer MoS$_2$ FETs with NGr electrodes [Fig. 4(d)]. We observed significant changes in the CNP of NGr (-49.2 V to -9 V) and the value of $V_{th}$ of the monolayer MoS$_2$ FETs (-54.2 V to -18.8 V) before and after thermal annealing in vacuum. This shows that electron doping due to nitrogen desorption in NGr changes the $V_{th}$ of MoS$_2$ with NGr after annealing. The field-effect mobility of the device after thermal annealing decreased from 8.76 cm$^2$/Vs to 1.02 cm$^2$/Vs, and the value of $V_{th}$ of the monolayer MoS$_2$ FETs with NGr electrodes after annealing shifted by 35.4 V in the positive direction. The mobility of the MoS$_2$ FETs appeared to have been influenced by changes in the electrical properties of NGr, rather than from the degradation of MoS$_2$ after thermal annealing.

In this letter, we reported high-performance monolayer MoS$_2$ FETs, where large-scale CVD-grown NGr graphene with an intrinsic high electron carrier density was used to establish Ohmic contact with monolayer MoS$_2$ at a modest gate voltage. The fabricated MoS$_2$ FETs exhibited a large negative shift in threshold voltage and barrier-free Ohmic contact at zero gate voltage owing to the Fermi level alignment of the monolayer MoS$_2$ and NGr. Moreover, the use of NGr in case of contact led to remarkable enhancements in the on-current and electron mobility of the monolayer MoS$_2$ FETs compared with when the pristine graphene electrode platform was used. Further, the atomic C-N bonding in NGr was responsible for the dominant effects of electron doping. We proposed a strategy that allows for the large-scale production of NGr electrodes as a promising device platform for the development of high-performance electronic devices, and for the examination of unique spin-valley physics for future electronics based on 2D materials.

## Supplementary material

See the supplementary materials for fabrication, the mobility of NGr, and the Raman and PL spectroscopies of the $MoS_2$ and NGr studied here after annealing.

## ACKNOWLEDGMENTS

D.S. was supported by the Graduate School of YONSEI University's research scholarship grants in 2017 (NRF-2018M3D1A1058924). D.S. and H.J.C. were supported by the Creative Materials Discovery Program through the National Research Foundation of Korea (NRF), funded by the Ministry of Science and ICT (2018M3D1A1058536). K.S.K. was supported by the Priority Research Center Program (2010-0020207) of the National Research Foundation (NRF) of Korea, funded by the Ministry of Education, and the Global Research & Development Center Program (2018K1A4A3A01064272) of the NRF funded by the Ministry of Science and ICT. J.H. was supported by the NSF MRSEC program through Columbia in the center of Precision Assembly of Superstratic and Superatomic Solid (DMR-1420634). G.-H.L was supported by Basic Science Research Program through the National Research Foundation of Korea (NRF) funded by the Ministry of Science, ICT & Future Planning (NRF-2017R1A2B2006568) and the Korea Institute of Energy Technology Evaluation and Planning (KETEP) and the Ministry of Trade, Industry & Energy (MOTIE) of the Republic of Korea (No. 20173010013340). Y.D.K. and J.J.L. were supported by Samsung Research & Incubation Funding Center of Samsung Electronics under Project Number SRFC-TB1803-04. This work was supported by a grant from Kyung Hee University in 2017 (KHU-20171743).

## Author contributions

D.S., J.K., and J.J.L. conceived of the experiment. D.Y.L. and K.S.K. grew the nitrogen-doped graphene. T.T. and K.W. grew the crystals of hexagonal boron nitride. J.H. and Y.D.K. advised on the experiments. D.S., H-J.C, and Y.D.K wrote the manuscript with K.S.K and G-H.L. All authors contributed to the overall scientific interpretation and editing of the manuscript.

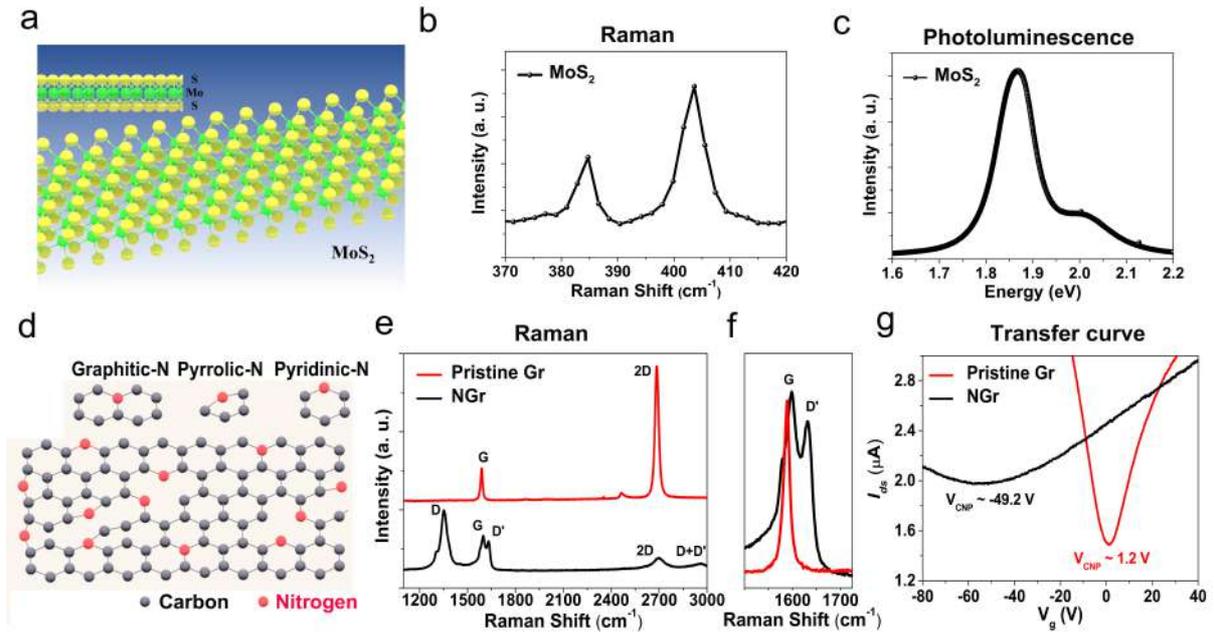

FIG. 1. (a) Schematic image of atomic structure of MoS$_2$. Raman (b) and PL (c) spectra of exfoliated MoS$_2$ at a laser excitation wavelength of 532 nm. (d) Schematic of three types of C-N atomic configurations in NGr: graphitic, pyrrolic, and pyridinic-N configurations. (e) and (f) Raman spectra of NGr and pristine graphene at a laser excitation wavelength of 532 nm. (g) Transfer curve of NGr and pristine graphene measured at constant drain voltage (10 mV) in vacuum. The charge neutral points of NGr (-49.2 V) and pristine graphene (1.2 V), respectively.

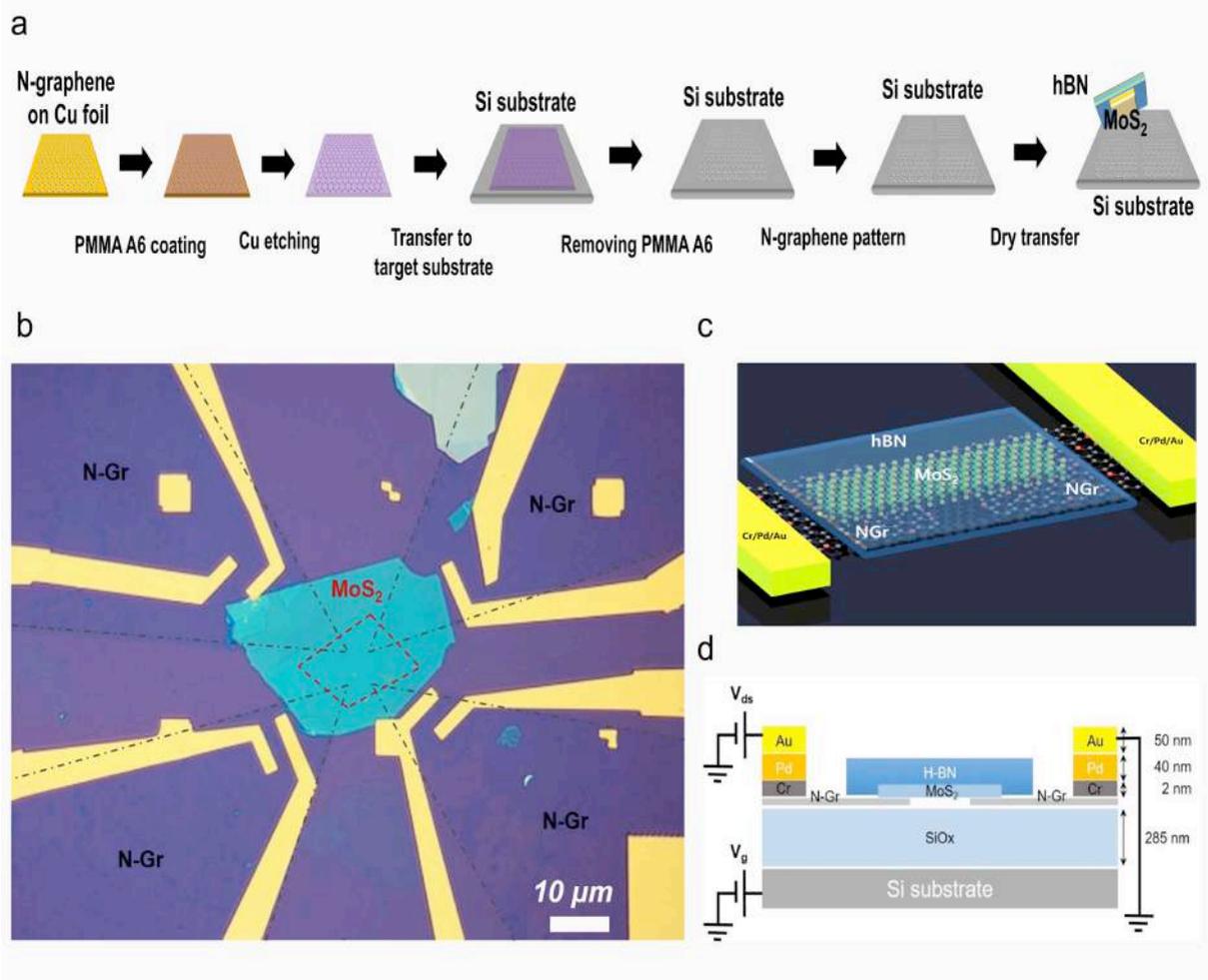

FIG. 2. Fabrication of MoS$_2$ FETs device with NGr electrodes. (a) Schematic diagrams of liquid transfer of CVD N-doped graphene, and its patterning, dry transfer, and device fabrication. (b) Microscopic image of MoS$_2$ FETs with a top hBN on NGr electrodes. (c) Schematic image of device structure of MoS$_2$ FETs with NGr electrodes. (d) Schematic of MoS$_2$ FETs with NGr electrodes showing the electrical connections. The channel MoS$_2$ was covered on top with hBN, and the 285nm SiO$_2$ served as the gate dielectric.

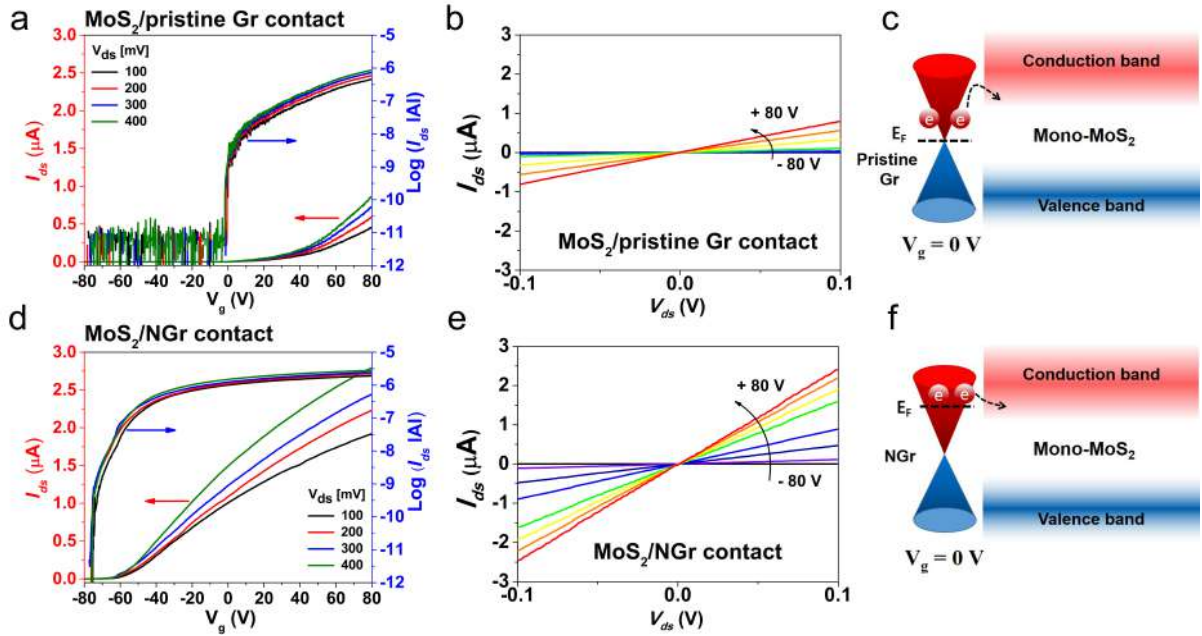

FIG. 3. Electrical characteristics of MoS$_2$ FETs in contact with pristine graphene and NGr. MoS$_2$ FETs with pristine graphene: (a) Transfer curve in linear (left) and semi-logarithmic scales (right), with $V_{ds}$ ranging from 100 mV to 400 mV in steps of 100 mV. The transistor shows a threshold voltage of $V_{th}$ = 39 V and on/off ratio of 10$^5$. (b) The output curve sweeps from -0.1 to 0.1 V with varying $V_g$ from -80 V to +80 V. (c) Schematic image of band alignment of MoS$_2$ FETs with pristine graphene at $V_g$ = 0 V. MoS$_2$ FETs with NGr: (d) Transfer curve in linear (left) and semi-logarithmic scales (right), with $V_{ds}$ ranging from 100 mV to 400 mV in steps of 100 mV. The transistor shows a threshold voltage of $V_{th}$ = -59.2 V and on/off ratio of 10$^6$. (e) The output curve sweeps from -0.1 to 0.1 V with varying $V_g$ from -80 V to +80 V. (f) Schematic image of band alignment of MoS$_2$ FETs with NGr at $V_g$ = 0 V.

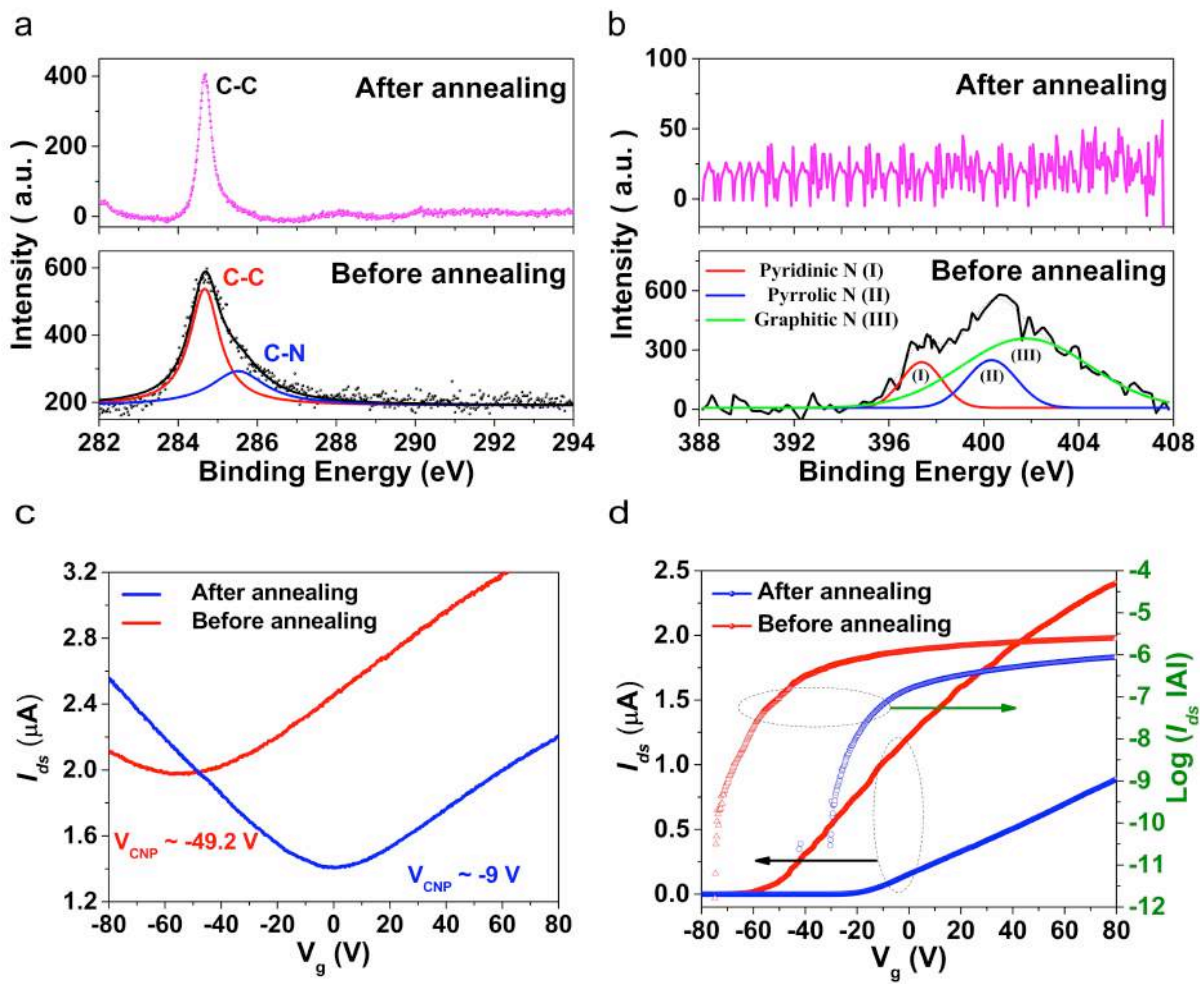

FIG. 4. Analysis of NGr after thermal annealing in vacuum at 360 °C for 30 min. High-resolution C 1s (a) and N 1s (b) XPS spectra showing bonding states in NGr before and after thermal annealing in vacuum, including the formation of pyridinic (I), pyrrolic (II), and graphitic (III) N structures. (c) Transfer curve of NGr before and after annealing. (d) Transfer curve of MoS$_2$ FETs with NGr contact before and after thermal annealing in vacuum. The threshold voltage ($V_{th}$) of MoS$_2$ FETs with NGr contact was changed from –54.2 V (before annealing) to -18.8 V (after annealing).

# Supporting Information

## S1. Fabrication of MoS$_2$ FET with NGr electrodes

### Synthesis of NGr

Common methods for replacing carbon by substitutional nitrogen in the structure of graphene are used to form chemical-vapor-deposited (CVD) nitrogen-doped graphene (NGr) with low defect densities and high performance. The mixed flow of carbon and nitrogen precursors in the gas phase causes competition between nitrogen incorporation and the NH$_x$ group-based etching effects that increase defect density.[1] The carbon- and nitrogen-mixed gas causes several problems in case of the N-C configuration, such as insufficient doping of graphene with defect density in the lattice and poor device performance. A low defect density in NGr substantially attenuates the induced short-range scattering. However, the method of synthesizing NGr without transport degradation has limitations. To overcome the bottleneck in the direct synthesis of NGr by the conventional CVD process and obtain good transport properties, the NGr was synthesized by CVD with pyridine (C$_5$H$_5$N), which is a nitrogen-containing organic precursor.[2] The liquid pyridine was added to a test tube (small chamber) connected to the main chamber of the CVD equipment, and pyridine was supplied by the difference of pressure by controlling the gauge valve between the test tube and the main chamber of CVD. After loading copper foil into the CVD chamber, it was heated to 1000 °C for 40 min with hydrogen (50 sccm) and Ar (100 sccm) gases at ambient pressure. When the temperature reached 1000 °C, the supply of Ar gas was shut off and the copper foil was annealed for 20 min under hydrogen flow (50 sccm). The supply of hydrogen gas was then shut off, and the gauge valve of the pyridine supply line was adjusted to supply vaporized pyridine. At this point, the pressure of the reaction chamber was $1.0 \times 10^{-1}$ Torr, and the NGr was synthesized for only 1 min. After synthesis, the supply valve of the pyridine source was shut off, the furnace was turned off, and the product was cooled to room temperature by passing Ar gas at 100 sccm through it.

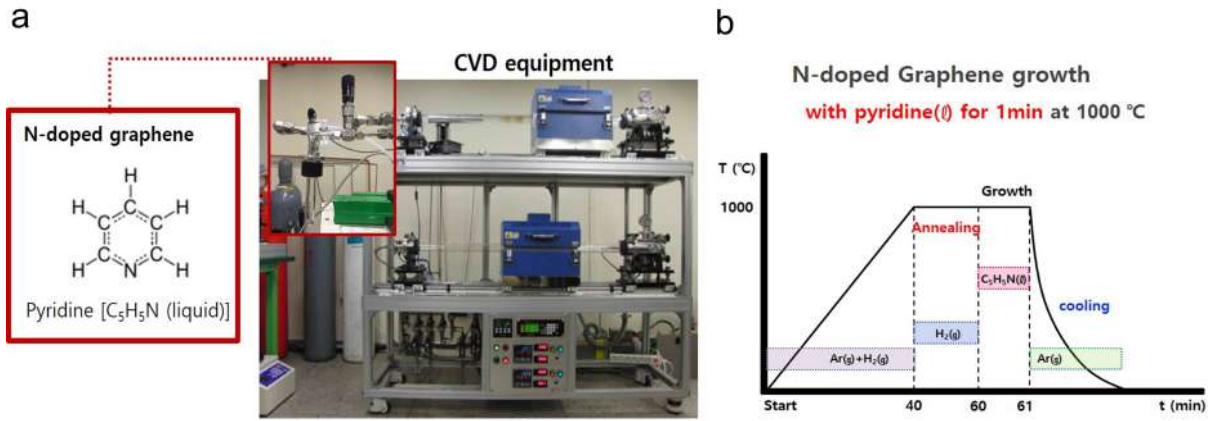

FIG. S1. CVD system (a) and synthesis process (b) of NGr with liquid pyridine ($C_5H_5N$) organic precursor.

**Fabrication of $MoS_2$ FETs**

We employed a liquid transfer process using a polymer to transfer the CVD-grown NGr to a target substrate. In this process, poly(methyl methacrylate) (PMMA; 950 A6, MicroChem) was coated on the CVD-grown NGr on Cu and baked at 180 °C for 5 min on a hot plate. The PMMA/Gr/Cu sample was then placed in a copper etchant, a 1 M ammonium persulfate solution (($NH_4)_2S_2O_8$, Sigma Aldrich), for 6 h and the $(NH_4)_2S_2O_8$ residue was removed by washing in a deionized water bath three times for 2 h. The sample was then scooped onto a $Si/SiO_2$ (285 nm) substrate by approaching the sample perpendicularly to the substrate. To improve the adhesion of NGr and prevent a bubble trap, the sample was kept in vacuum for about one day. It was then placed in acetone to remove PMMA (950 A6). NGr was patterned by e-beam lithography using hydrogen silsesquioxane (HSQ) and etched with $O_2$ plasma. We manufactured $hBN/MoS_2/NGr$ using a dry pick-up technique. We used a polypropylene carbonate (PPC)-coated poly(dimethylsiloxane) (PDMS) block mounted on a glass slide for pick-up and release.[3,4] We first picked up the exfoliated hBN and then the $MoS_2$ to fabricate the $hBN/MoS_2$ heterostructure. Then, the $hBN/MoS_2$ heterostructure was transferred onto the patterned NGr electrode. To obtain clean $MoS_2$ free of the polymer residue, the $hBN/MoS_2$ heterostructure was transferred onto the patterned NGr electrode. The surface of $MoS_2$ was covered entirely by hBN to prevent contamination by external polymer residues. Metal leads, Cr (2 nm)/Pd (40 nm)/Au (50 nm), were directly deposited on the NGr electrodes by e-beam evaporation. For contact between NGr and the monolayer $MoS_2$, only the NGr region was metalized.

## S2. Raman spectroscopy of CVD-grown NGr

In the NGr (black line in Fig. 1(e)), a relatively low intensity of the 2D peak (~2702.12 cm$^{-1}$) compared with that of pristine graphene was observed due to nitrogen-induced electron doping. The slightly blue-shifted G peak (~1596.68 cm$^{-1}$) of NGr compared with the G peak (~1588.52 cm$^{-1}$) of pristine grapheme was also noted. There are several reasons for the shift in the Raman peak, including the effects of both doping and strain.[5] This occurred due to the addition of nitrogen as an impurity into the graphene lattice, which resulted in a phonon oscillation mode associated with the defect. The G-peak shifted and the intensity of the 2D peak weakened owing to distorted lattice spacing, and different masses and charge distributions, which provide indirect evidence of the effect of the doping of nitrogen. The intensities of both the D and the D′ peaks increased with an increase in the concentration of the defects. Nitrogen dopants in the graphene lattice created defects in it by electron doping. Axel et al.[6] have reported that the intensity ratio of $I_D/I_{D'}$ can be used experimentally to obtain information on the nature of defects in graphene by analyzing the Raman spectra of graphene containing different types of defects. The characteristics of defects in graphene were determined by the $I_D/I_{D'}$ intensity ratio; $I_D/I_{D'}$ was ~13 for sp$^3$ defects, ~7 for vacancy-type defects, and ~3.5 for the domain boundary. In our CVD-grown NGr, the $I_D/I_{D'}$ intensity ratio was ~1.6, close to the value of the domain boundary defects. This is related to both the lattice distortion of pyrrolic, pyridinic, and graphitic N configurations, and the grain boundary.

## S3. Electrical characterization and mobility of NGr

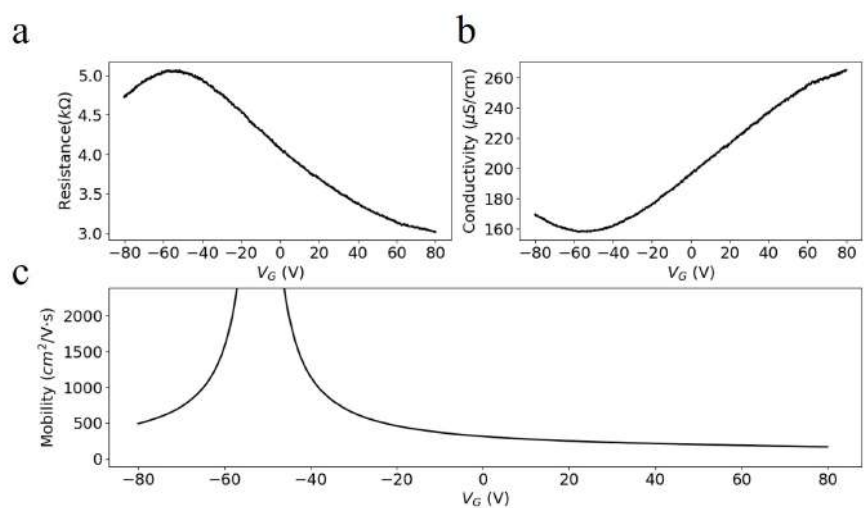

FIG. S2. Electrical characteristics of NGr: resistance (a), conductivity (b), and mobility (c) versus gate voltage.

## S4. Electrical characterization

We carried out the electrical characterization of the device using an SR830 lock-in amplifier and a Keithley 2400 system. All electrical properties were measured at room temperature and low temperature (77 K) in vacuum. We performed electrical transport measurements under a constant source–drain voltage of 10 mV and observed the charge neutrality point (CNP) of pristine graphene and NGr. To determine the two-probe transfer characteristic of MoS$_2$ FET, a constant drain voltage ($V_{ds}$) was applied from 100 mV to 400 mV in steps of 100 mV. The Si/SiO$_2$ (285 nm) substrate was used as a global back gate, where the bias voltage was used to control the carrier concentrations of pristine graphene and NGr.

For MoS$_2$ with pristine graphene, the threshold voltage was between +20 V and +40 V. When NGr was used as an electrode for MoS$_2$, $V_{th}$ was between -45 V and -55 V. We observed the difference in terms of $V_{th}$ and FET mobility between the MoS$_2$ with pristine graphene and electron-doped NGr, respectively. We confirmed a similar tendency by measuring several samples.

| MoS$_2$ | MoS$_2$ FETs ||||||||
|---|---|---|---|---|---|---|---|---|
| | With pristine graphene |||| With NGr ||||
| | *Device* | Gr - $V_{CNP}$ | MoS$_2$ - $V_{th}$ | Mobility (cm$^2$/V-s) | *Device* | NGr - $V_{CNP}$ | MoS$_2$ - $V_{th}$ | Mobility (cm$^2$/V-s) |
| **Monolayer** | # 1 | 1.2 V | 39 V | 2.21 | # 1 | -49.2 V | -54.2 V | 8.76 |
| **Monolayer** | # 2 | 5 V | 28.2 V | 1.98 | # 2 | -38.1 V | -50.8 V | 5.78 |
| **Monolayer** | # 3 | 7 V | 20.7 V | 3.01 | # 3 | - 20.5 V | - 45 V | 6.01 |

**Table S1**. Electrical properties of MoS$_2$ FETs with pristine graphene and NGr, respectively.

In addition, we demonstrated the MoS$_2$ device with different numbers of layers, from a monolayer to a multi-layers, to check the electrical properties of MoS$_2$ with NGr before and after annealing. After annealing, the values of $V_{th}$ of most MoS$_2$ devices shifted in the positive direction and the mobility of MoS$_2$ decreased despite insignificant changes to its Raman and PL.

| | MoS$_2$ | NGr ($V_{CNP}$) | MoS$_2$ FETs with NGr contact ||||
|---|---|---|---|---|---|---|
| | | | Before annealing || After annealing ||
| | | | $V_{th}$ | Mobility (cm$^2$/V-s) | $V_{th}$ | Mobility (cm$^2$/V-s) |
| # 1 | **Monolayer** | -49.2 V | -54.2 V | 8.76 | -18.8 V | 1.02 |
| # 2 | **Monolayer** | -38.1 V | -50. 8 V | 5.78 | -28 V | 2.98 |
| # 3 | **Monolayer** | 20.5 V | -45 V | 6.01 | -35 V | 2.77 |
| # 4 | **Bilayer** | -21 V | -60 V | 10.63 | -30 V | 5.11 |
| # 5 | **Bilayer** | -41.2 V | -55.2 V | 7.52 | -25 V | 2.09 |
| # 6 | **Multi-layers (9 nm)** | -39.5 V | -57 V | 41.1 | -22 V | 10.4 |

**Table S2**. Electrical properties of different number of $MoS_2$ layers from monolayer to multi-layers with NGr before and after annealing.

## S5. *I-V* curve of $MoS_2$ with NGr and pristine graphene at 77 K

$MoS_2$ with NGr was clearly linear in the gate range from -80 V to +80 V at 77 K, whereas in case of pristine graphene, the contact showed high resistance and a non-linear output curve even at large gate voltage ($V_g$ = +100 V), indicating that the contacts were non-Ohmic.

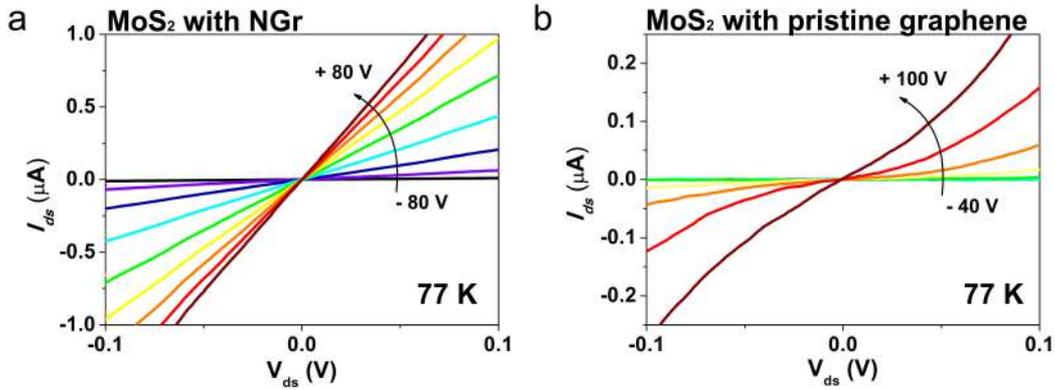

FIG. S3. (a) Linear output curve of $MoS_2$ with NGr with varying $V_g$ from -80 V to +80 V at 77 K. (b) Non-linear output curve of $MoS_2$ with pristine graphene K with varying $V_g$ from -40 V to +100 V at 77 K.

## S6. Raman and PL of $MoS_2$ with NGr electrodes before and after annealing

The Raman peaks of NGr appeared at the same position, although the intensity of the D´ peak decreased. The $I_D/I_{D'}$ intensity ratio of NGr was approximately ~1.16, close to the value of defects in the domain boundary, and was lower than that before annealing. This is related to the different defects in NGr before and after annealing.[7] Moreover, a vacancy defect might also have resulted from the desorption of substitutional nitrogen dopant in NGr.

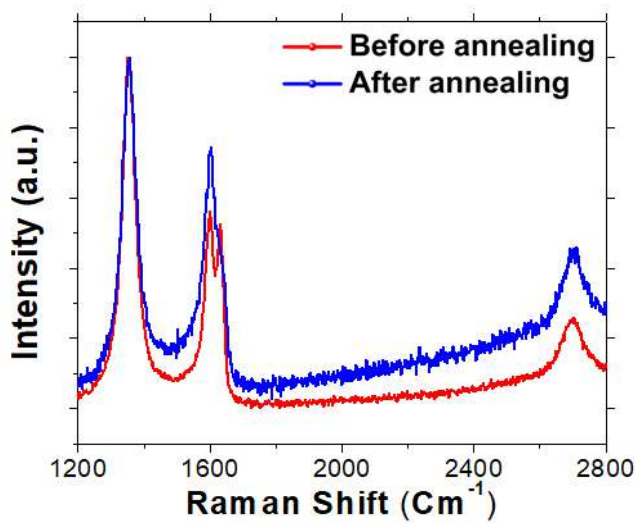

FIG. S4. Raman spectra of NGr before and after thermal annealing in vacuum.

The shift and changes in intensity in the Raman peak of $MoS_2$, an in-plan ($E_{2g}$) mode at 384.72 cm$^{-1}$ and an out-of-plane ($A_{1g}$) mode at 403.62 cm$^{-1}$, were almost negligibly small. The PL of the representative $MoS_2$ before and after annealing was normalized to the maximum intensity of emission. The PL of $MoS_2$ showed a red shift of 0.02 eV at the peak after annealing, where the change was not significant. The change in the PL peak is thought to be the effect of strain.[8]

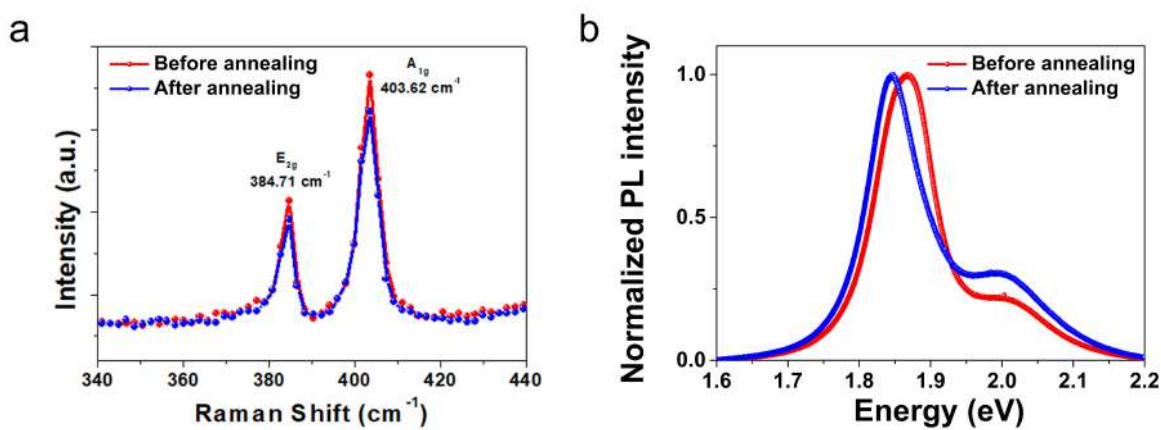

FIG. S5. Raman spectra (a) and PL spectra (b) of $MoS_2$ before and after thermal annealing in vacuum.

All spectra exhibited emissions from both A and B excitons. The intensity for each peak was determined by Lorentz fitting. The value of the full-width-at-half-maximum (FWHM) of PL was related to the quality of the crystalline material. We analyzed the PL based on the measurement of the FWHM in each of its spectra. The sharp PL peak with a FWHM of 93 meV (before annealing) and 86 meV (after annealing) of the A-exciton also indicates that hBN-covered $MoS_2$ remained its crystal structure without degradations.

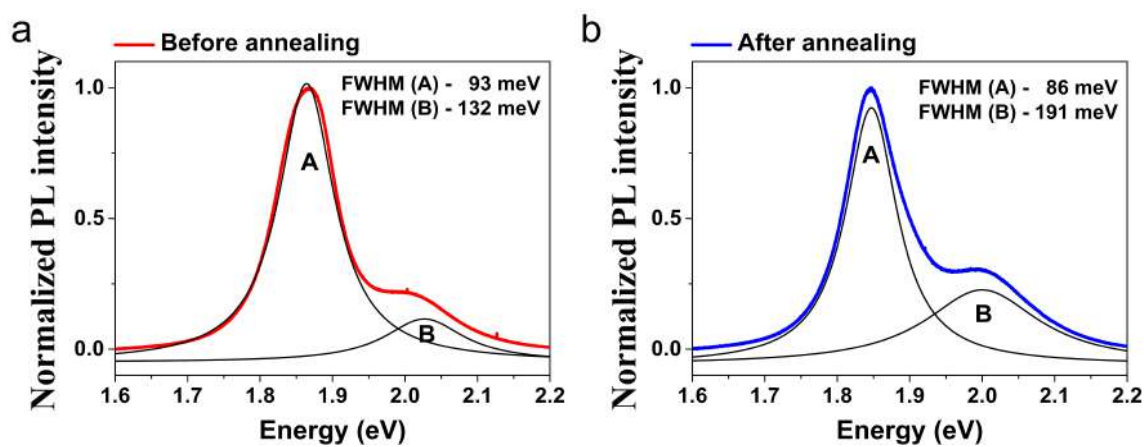

FIG. S6. The value of full-width-at-half-maximum (FWHM) of PL before (a) and after (b) annealing.